\documentstyle[proceedings,numreferences,epsfig]{tasy}

\begin{opening}
\title{Is a quantum theory of resonances really time asymmetric?}
\author{F. Kleefeld}
\institute{Centro de F\'{\i}sica das Interac\c{c}\~{o}es Fundamentais (CFIF),\\
Instituto Superior T\'{e}cnico,\\ 
Av. Rovisco Pais, P-1049-001 LISBOA, Portugal\\  
{\sf kleefeld@cfif.ist.utl.pt,
 http://cfif.ist.utl.pt/$\sim$kleefeld/}}
\end{opening}
\runningtitle{Is a QT of resonances really time asymmetric?}
\begin{document}
\begin{abstract}
The title ``Time Asymmetric Quantum Theory: the Theory of Resonances'' (without questionmark) of the CFIF workshop (23.--26.7.2003, Lisbon, Portugal) implies that the theoretical description of resonances is uniquely described by the formalism of A.\ Bohm et al.\ \cite{Bohm:2002vt,Bohm:2002ja} reflecting the title of the workshop. Our presentation in this workshop tries to introduce an apparently inequivalent, alternative feasible relativistic formalism provided by the author \cite{Kleefeld:2003zj,Kleefeld:2002au,Kleefeld:2002gw,Kleefeld:2001xd,Kleefeld:1998yj,Kleefeld:1998dg,Kleefeld:thesis1999} under the name ``(Anti)Causal Quantum Theory'' ((A)CQT) which is compared to the former. 
\end{abstract}

\section{The ``Time Asymmetric Quantum Theory'' of A.\ Bohm et al.}
One longstanding debate on Quantum Theory (QT)
finds its origin in a logical discrepancy between the description of stationary and time-dependent quantum problems. To put it in the words of A.~Bohm et al.\ \cite{Bohm:2002vt}:\\ 
{\em ``$\ldots$ Quantum Theory in Hilbert space is time symmetric. This is not so bad for the description of spectra and structure of quantum physical systems, whose states are (or are considered as) stationary. But this is particularly detrimental for the description of decay processes and resonance scattering, which are intrinsically irreversible processes. There is no consistent theoretical description for decaying states and resonances in Hilbert space quantum mechanics \footnote{Note on the contrary P.T.\ Matthews \& A.\ Salam \cite{Matthews:1958sc,salam1958} or N.\ Nakanishi \cite{nakanishi1958}!} $\ldots$''} In order to overcome this problem A.\ Bohm et al.\ suggest to change one axiom of Quantum Mechanics (QM).\footnote{To use their words \cite{Bohm:2002vt}:\\ 
{\em ``$\ldots$ Whereas the standard Hilbert space axiom does not distinguish mathematically between the space of states (in-states of scattering theory) and the space of observables (out-`states' of scattering theory) the new axiom associates states and observables to two different Hardy subspaces} [$\Phi_+$ and $\Phi_-$] {\em which are dense in the same Hilbert space} [${\cal H}$] {\em and analytic in the lower and upper complex energy plane, respectively $\ldots$''} 

Denoting $\Phi^\times$ to be the space of continuous antilinear functionals on $\Phi$ they state \cite{Bohm:2002vt}:\\ 
{\em ``$\ldots$ The solutions of the dynamical equation $i\partial_t \psi^+(t) = H \psi^+(t)$ with the new boundary condition $\{\phi^+\}=\Phi_-\subset {\cal H} \subset \Phi^\times_-$ are for the states $\phi^+\in \Phi_-$ (i.e.\ the set of states defined physically by the preparation appatuses (accelerator)), given by $\phi^+(t)=\exp(-i H t)\,\phi^+$; $0\le t<\infty$. The solutions of the dynamical equation $i\partial_t \psi^-(t) = - \, H \psi^-(t)$ with the new boundary condition $\{\phi^-\}=\Phi_+\subset {\cal H} \subset \Phi^\times_+$ are for the observables $\phi^-\in \Phi_+$ (i.e.\ the set of states the set of states defined by registration appatuses (detector)), given by $\phi^-(t)=\exp(i H t)\,\phi^-$; $0\le t<\infty$. $\ldots$''} 

According to A.\ Bohm et al.\ \cite{Bohm:2002vt} the old axioms of QM yield a {\em ``reversible time evolution $\ldots$ given by the unitary group $U(t)=\exp(i H t)$} ({\em or} $U^+(t)=\exp(-i H t)$) {\em with $-\infty<t<\infty$''}, while the new axioms lead only to {\em ``a semigroup time evolution $0\le t<\infty$, which cannot be reversed to negative time. This singles out a particular time $t_0$, the mathematical semigroup time $t_0=0$. $\ldots$ We interpret this $t_0$ as the time at which the state has been prepared and at which the registration of an observable in this state can start. $\ldots$ It is in particular the choice of the Hardy spaces for the [new] axiom $\ldots$ that leads to time asymmetry.''}. 

Finally \cite{{Bohm:2002ja}} they extend their space of states by Gamow (and anti-Gamov) vectors with complex energy eigenvalues, in order to describe resonances, and introduce the concept of a Causal Poincar\'{e} Semigroup:\\
{\em ``$\ldots$ The Poincar\'{e}  transformations of the Gamow kets, as well as of the Lippmann-Schwinger plane wave scattering states, form only a semigroup of Poincar\'{e}  transformations  into the forward light cone $\ldots$''}} 
We note that A.\ Bohm et al.\ pay a severe price to preserve Born's probability interpretation \cite{Pais:1982we} as an axiom of QM: the future and the past is separated by a ``mathematical semigroup time'' $t_0$ which is hardly reconcilable with the requirement translational invariance in time even in the absence of resonances. Further important questions arise: How about causality and four-momentum conservation when propagating from $t<t_0$ to $t>t_0$ or when asymptotic states interact through intermediate resonances? Why are only resonances/unstable particles treated like (anti)Gamow states with complex energy eigenvalues while other states are associated in the traditional manner with real eigenvalues? How is the formalism generalized to Quantum Field Theory (QFT)? How about boost properties of spinors and polarization vectors of resonances/unstable particles described by (anti)Gamow vectors? How about aspects like gauge invariance in the presence of resonances which are relevant in describing e.g.\ the excitation of intermediate vector-Bosons like $Z^0$ or $W^\pm$? 

\section{Introduction to (Anti)Causal Quantum Field Theory}

Considering the Schr\"odinger equation as a particular limit of the Klein-Gordon (KG) equation it is advantageous to introduce first the formalism of (A)CQT within the context of relativistic QFT theory and then to consider its limit to Schr\"odinger's representation of QM. With respect to space constraints we keep the presentation rather short and refer for details to previous work (see e.g.\ Refs.\ \cite{Kleefeld:2003zj,Kleefeld:2002au,Kleefeld:2002gw,Kleefeld:2001xd,Kleefeld:1998yj,Kleefeld:1998dg,Kleefeld:thesis1999}).

The key to (A)CQT providing answers to questions raised in the previous section is the following conjecture by P.B.\ Burt (p.\ 29 in \cite{Burt:bp}): 

{\em ``$\ldots$ Already on the classical level the most general physics is obtained
by solving the equations of motion without constraint. In fact, imposing
constraints prior to the solution of the equations of motion can lead one to
erroneous conclusions. In parallel, in quantum field theory we adopt the premise that non-Hermitian solutions of the field equations are acceptable. The physics contained in the hermiticity assumption will be built in after explicit
solutions are constructed. $\ldots$''}

What does this mean in detail? Textbooks (e.g.\ \cite{Kleefeld:Bjorken:dk}) on QFT declare the neutral scalar KG field to be a {\em Hermitian} ({\em shadow} \cite{Kleefeld:Nakanishi:wx,Kleefeld:Stapp:1973aa}) {\em field} $\varphi(x)=\varphi^+(x)$ with real mass $m=m^\ast$ and Lagrangean ${\cal L}^{\,0}_{\,\varphi} (x) = \frac{1}{2}\left( (\partial\varphi (x) )^2  - \, m^2 \, \varphi (x)^2 \right)$. Note that $\varphi(x)$ is representing {\em one} (real) field-theoretical degree of freedom! Its equation equation of motion is given by \mbox{$(\partial^2 + m^2)\, \varphi (x) = 0$} yielding --- strictly speaking --- a principal value propagator \cite{Kleefeld:Stapp:1973aa} $\mbox{P} \,\frac{i}{p^2-m^2}$. Causality is typically enforced {\em afterwards} by the use of {\em causal} or {\em anticausal} Feynman propagators ($i/(p^2-m^2+i\,\varepsilon)$ (causal), $i/(p^2-m^2-i\,\varepsilon)$ (anticausal)) corresponding --- strictly speaking --- to the {\em causal KG equation} $(\partial^2 + m^2 - i\,\varepsilon )\, \phi (x)\, = 0$ or {\em anticausal KG equation} $(\partial^2 + m^2 + i\,\varepsilon )\, \phi^+(x) = 0$, respectively \cite{Kleefeld:Nakanishi:wx,Kleefeld:Nakanishi:1972pt,Kleefeld:2002au,Kleefeld:2002gw,Kleefeld:2001xd}.
The {\em causal KG field} $\phi(x)=(\varphi_1(x)+i\,\varphi_2(x))/\sqrt{2}$ and the {\em anticausal KG field} $\phi^+(x)=(\varphi_1(x)-i\,\varphi_2(x))/\sqrt{2}$ are {\em non-Hermitian}\footnote{The non-Hermitian nature of QT should not surprise, but be taken into account! In-fields and out-fields fulfil same {\em causal} KG equations: $(\partial^2 + m^2 - i\,\varepsilon ) \, \phi_{in} (x) = 0$, $(\partial^2 + m^2 - i\,\varepsilon ) \,\phi_{out} (x) = 0$. Hence, the space of out-states is not obtained from the space of in-states by Hermitian conjugation. Therefore we claim that in QM ``bra's'' (in our notation: $\left<\!\left<\ldots\right|\right.\!$) are not obtained by Hermitian conjugation from ``ket's'' (in our notation: $\left|\ldots\right> = \left<\ldots \right|^+$)!} and represented by {\em two Hermitian shadow fields}  $\varphi_j(x)=\varphi_j^+(x)$ ($j=1,2$) yielding {\em two} (real) field-theoretical degrees of freedom. I.e.\ imposing causal boundary conditions on QT leads (already at zero temperature) to a {\em doubling} of degrees of freedom like in Thermal Field Theory \cite{Kleefeld:Weldon:1998yk} or Open Quantum Systems \cite{Kleefeld:Romano:2003ms}. 

Most probably inspired by work of T.D.\ Lee and C.G.\ Wick \cite{Lee:iw} N.~Nakanishi \cite{Kleefeld:Nakanishi:wx,Kleefeld:Nakanishi:1972pt} investigated 1972 a Lagrangian for a KG field $\phi (x)$ with complex mass\footnote{In early works complex mass fields were discussed as e.g.\ ``dipole ghosts'' \cite{heis1957,heis1961},  ``complex roots'' \cite{pauli1958}, ``complex poles'' \cite{lee1970}, or ``complex ghosts'' \cite{Kleefeld:Nakanishi:wx}. Note also T.D.\ Lee, G.C.\ Wick \cite{Lee:fy,Lee:iw,Lee:ix}, A.M.\ Gleeson et al.\cite{Gleeson:sw,Gleeson:xj}, M.\ Froissart \cite{froi1959}, K.L.\ Nagy et al.\cite{nagy1970}.} $M := m - \frac{i}{2} \, \Gamma$ (and the Hermitian conjugate field $\phi^+(x)$)\footnote{The formalism$\;\!$was$\;\!$1999-2000 independently rederived by the author (see e.g.\ \cite{Kleefeld:2002au,Kleefeld:2002gw,Kleefeld:2001xd}).}. In the following we want to introduce immediately isospin and to consider for convenience a set of $N$ equal complex mass KG fields $\phi_r (x)$ ($r = 1,\ldots,N$) (i.e.\ a charged ``Nakanishi field'' with isospin $\frac{N-1}{2}$) described by the Larangean\\[1mm]
\makebox[3mm]{}${\cal L}_\phi^{\,0} (x) = \!\sum\limits_{r} \!\left\{
\frac{1}{2} \left( (\partial \phi_r (x) )^2  - M^2 \, \phi_r (x)^2 \right)\! + 
\frac{1}{2} \left( (\partial \phi_r^+ (x) )^2  - M^{\ast \, 2} \, \phi_r^+ (x)^2 \right) \right\}$.\\[1mm]
The Lagrange equations of motion for the causal and anticausal ``Nakanishi field'' $\phi_r (x)$ and $\phi_r^+ (x)$, i.e.\ $(\,\partial^2 + M^2) \,\phi_r (x) = 0$ and \mbox{$(\,\partial^2 + M^{\ast \, 2}) \,\phi_r^+ (x) = 0$},  are solved by a Laplace-transform. The result is:\\[1mm]
\makebox[5mm]{}$\phi_r (x) \, =  \int 
\frac{d^3 p}{(2\pi )^3 \; 2\, \omega \, (\vec{p}\,)}
\Big[ \, 
 a \, (\vec{p} , r ) \, e^{\displaystyle - \, i p x}  +
 c^+ (\vec{p} , r ) \, e^{\displaystyle i p x}
\,\Big] \Big|_{p^0 = \omega(\vec{p}\,)} \;$ ,\\[1mm]
\makebox[5mm]{}$\phi_r^+ (x) =  \int
\frac{d^3 p}{(2\pi )^3 2\omega^\ast (\vec{p}\,)}
\Big[ 
  c (\vec{p} , r ) \,e^{\displaystyle - i p^\ast x} +
 a^+ (\vec{p} , r ) \,e^{\displaystyle i p^\ast x}
\Big] \Big|_{p^0 = \omega(\vec{p})}\;$ , \\[2mm]
where we defined $a(\vec{p}\,):=a(p)|_{p^{0} = \omega (\vec{p}\,)}$ and $c^+(\vec{p}\,):=a(-p)|_{p^{0} = \omega (\vec{p}\,)}$ on the basis of the complex ``frequency'' $\omega(\vec{p}):=\sqrt{\vec{p}^{\,2}+M^{\,2}}$ ($\omega(\vec{0}):=M$)\footnote{To obtain this result we had to use a delta-distribution ``$\delta (p^2 - M^2)$'' for complex arguments which has been illuminated by N.\ Nakanishi \cite{Kleefeld:Nakanishi1958,Kleefeld:Nakanishi:1972pt}. Nowadays it may be embedded in the framework of (tempered) Ultradistributions \cite{Kleefeld:Bollini:1998en}.}. The ``Nakanishi model'' is quantized by claiming {\em Canonical equal-real-time commutation relations}. The resulting non-vanishing momentum-space commutation relations, which display an {\em indefinite metric}\footnote{An indefinite metric should not surprise, as the space-time metric is $(+,-,-,-)$!}, are ($r,s = 1,\ldots,N$):\\[1mm]
\makebox[5mm]{}$[ \, a \, (\vec{p},r) \; , \; c^+ (\vec{p}^{\,\,\prime},s) \, ]  = (2\pi)^3 \, 2 \; \omega \,(\vec{p}\,)\;\, \delta^{\, 3} (\vec{p} - \vec{p}^{\,\,\prime}\,) \; \delta_{rs} \;$ ,\\[1mm]
\makebox[5mm]{}$[ \, c \, (\vec{p},r) \; , \; a^+ (\vec{p}^{\,\,\prime},s) \, ] = (2\pi)^3 \, 2 \; \omega^\ast(\vec{p}\,)\; \delta^{\, 3} (\vec{p} - \vec{p}^{\,\,\prime}\,) \; \delta_{rs} \;$.\\[2mm]
The Hamilton operator is derived by a standard Legendre transform \cite{Kleefeld:2003zj,Kleefeld:2001xd}:\\[1mm]
\makebox[5mm]{}$H_\phi^{0} =
\sum\limits_{r} \;\int\! \!
\frac{d^3 p}{(2\pi )^3 \; 2\, \omega \, (\vec{p}\,)}
\;\;
\frac{1}{2}\;\, \omega \,(\vec{p}\,)\;\, \Big( 
c^+ (\vec{p},r) \; a (\vec{p},r) +
a\, (\vec{p},r) \; c^+ (\vec{p},r) \; \Big)$\\[1mm]
\makebox[11mm]{}$ + \, \sum\limits_{r} \,\;\int\! \!
\frac{d^3 p}{(2\pi )^3 \; 2\, \omega^\ast (\vec{p}\,)}
\;
\frac{1}{2}\; \omega^\ast (\vec{p}\,)\; \Big( 
a^+ (\vec{p},r) \; c\, (\vec{p},r) +
c \, (\vec{p},r) \; a^+ (\vec{p},r) \; \Big)$.\\[2mm]
The ``Nakanishi-KG propagator'' is obtained by {\em real-time ordering} of causal KG fields \cite{Kleefeld:Nakanishi:wx,Kleefeld:2001xd,Kleefeld:Bollini:1998hj}.\footnote{Explicitely we obtain:\\
\makebox[10mm]{}$\Delta_N (x-y) \;  \delta_{rs} := -\,i\, \left<\!\left<0\right|\right.T\,[\,\phi_r\,(x)\, \phi_s\, (y)\,]\,\left|0\right>  \stackrel{!}{=} \; {\displaystyle \int\!\frac{d^{\,4}p}{(2\,\pi)^4}\; \frac{e^{-\,i\,p (x-y)}}{p^2 - M^2} \; \delta_{rs}}\;$.\\[1mm]
\makebox[5.5mm]{}The anticausal ``Nakanishi-KG propagator'' is obtained by Hermitian conjugation or by a vacuum expectation value of an {\em anti-real-time ordered} product of two anticausal fields.
For intermediate states with complex mass these propagators lead to {\em Poincar\'{e} covariant} results. At each interaction vertex coupling to {\em intermediate} complex mass fields there holds {\em exact} 4-momentum conservation. Only if complex mass fields with finite $\Gamma$ appeared as asymptotic states, then Poincar\'{e} covariance would be violated!} 
It is instructive to decompose an (A)CQT into its Hermitian components. Hermitian fields underlying non-Hermitian (anti)causal fields are here called ``shadow fields'' \cite{Kleefeld:Nakanishi:wx,Kleefeld:Stapp:1973aa}\footnote{E.C.G.\ Sudarshan \cite{Gleeson:xj} used the term ``shadow state'' with a {\em slightly different} meaning!}. Consider e.g.\ the (anti)causal Lagrangean of a neutral (anti)causal spin 0 Boson:\\[1mm]
${\cal L}^{\,0}_{\,\phi} (x) =  
\frac{1}{2}\, \Big( (\partial \,\phi (x) )^2  - \, M^2 \, (\phi (x))^2 \, \Big)\; + \;
\frac{1}{2}\, \Big( (\partial \,\phi^+ (x) )^2  - \, M^{\ast \, 2} \,
(\phi^+ (x))^2 \Big)$.\\[1mm] 
$\phi(x)$, $\phi^+(x)$ are decomposed in Hermitian shadow fields $\phi_{(1)}(x)$, $\phi_{(2)}(x)$ by  $\phi(x) =: ( \phi_{(1)}(x) + i \, \phi_{(2)} (x))/\sqrt{2}$, $\phi^+(x) =: ( \phi_{(1)}(x) - i \, \phi_{(2)} (x))/\sqrt{2}$ yielding the decomposed Lagrangean\\[1mm]
\makebox[3mm]{}${\cal L}^{\,0}_{\,\phi} (x) = 
\frac{1}{2} \,\Big( (\partial \phi_{(1)} (x) )^2  - \mbox{Re}[M^2]  
(\phi_{(1)} (x))^2 \Big)$\\[2mm]
\makebox[14.2mm]{}$ - \; \frac{1}{2} \,\Big( (\partial \phi_{(2)} (x) )^2  -  \mbox{Re}[M^2]  (\phi_{(2)} (x))^2 \Big) 
 + \mbox{Im}[M^2]  \; \phi_{(1)} (x) \,\phi_{(2)} (x)  \;$.\\[1mm]
Note that shadow fields are described by {\em principal value propagators} and {\em interact} with each other. One shadow field has {\em positive} norm, one has {\em negative} norm. If one would remove the interaction term, one would introduce interactions between causal and anticausal fields (e.g. $\phi(x) \, \phi^+(x)$) leading to a {\em violation of causality, analyticity and locality} in QT.\footnote{N.\ Nakanishi \cite{Nakanishi:jj,Kleefeld:Nakanishi:wx} and A.M.\ Gleeson et al.\ \cite{Gleeson:xj} demonstrate in this case even a {\em Lorentz noninvariance} of the S-matrix. Up to now most models studied complex mass fields in the presence of real mass asymptotic states (naively describing the so called ``physical Hilbert space'') or allowed interactions between causal and anticausal fields and ran therefore into various of these problems (e.g.\ \cite{pauli1958,lee1970,Gleeson:sw,Gleeson:xj,nagy1970,Nakanishi:jj,Kleefeld:Nakanishi:wx}). A further example will be the ``Time Asymetric Quantum Theory'' of A.\ Bohm et al.\ \cite{Bohm:2002vt,Bohm:2002ja}, if they continue to consider {\em only} resonances/unstable particles as (anti)Gamow states!} The formalism of the (anti)causal KG fields can be extended \cite{Kleefeld:2003zj,Kleefeld:2002gw} to (anti)causal Dirac fields and (anti)causal (non)Abelian vector fields leading to gauge-invariant, causal, local and Lorentz invariant (non)Abelian gauge theories, if {\em all} fields (including fields representing asymptotic states) are treated non-Hermitian and {\em no} interactions between causal and (anti)causal fields are allowed.\footnote{I.e.\ the formalism of (A)CQT is based on the idea that (anti)causal fields are fields with a complex mass containing at least a non-vanishing infinitesimal imaginary part. Hence, such (anti)causal fields are not allowed to be treated Hermitian as suggested in all present textbooks. Even if the overall Lagrangean (and Hamiltonian) of (A)CQT is --- for unitarity reasons --- Hermitian, it consists of a sum of two non-Hermitian terms. One term (i.e.\ the so-called causal Lagrangean) describes the causal propagation of causal fields (towards the future), the other term (i.e.\ the so-called anticausal Lagrangean) is Hermitian conjugate to the former and describes the anticausal propagation of anticausal fields (towards the past). We happen to live in a world which is described by the causal Lagrangean, while an enventual time-reversed universe described by the anticausal Lagrangean seems to be out of our observability (if we disregard gravitation as a possible source of information). As the causal and anticausal description is related by Hermitian conjugation the formalism of (A)CQT is not really time asymmetric.}
For completeness we show here also the Lagrangean of the (anti)causal Dirac field (isospin index $\;r=1, \ldots, N$) ($M := m - \frac{i}{2} \, \Gamma$, $\bar{M}:= \gamma_0 \, M^+ \gamma_0$)\footnote{The Lagrangian for $N=2$ was for the first time denoted by T.D.\ Lee and C.G.\ Wick \cite{Lee:iw} (see also I.I.\ Cot\v{a}escu \cite{Cotaescu:nc}), and then later independently rederived for arbitrary integer $N$ by the author \cite{Kleefeld:2003zj,Kleefeld:2002au,Kleefeld:2002gw,Kleefeld:1998yj,Kleefeld:1998dg,Kleefeld:thesis1999}.}
\makebox[3mm]{}$ {\cal L}^{\,0}_{\,\psi} (x) = \sum\limits_r \,\frac{1}{2} \Big( \; \overline{\psi_r^c} (x) \, ( \frac{1}{2} \, i \! \stackrel{\;\,\leftrightarrow}{\partial\!\!\!/} \! -  M ) \, \psi_r (x) \; + \; \bar{\psi}_r (x) \, ( \frac{1}{2} \, i \! \stackrel{\;\,\leftrightarrow}{\partial\!\!\!/} \! -  \bar{M} ) \, \psi_r^c (x) \; \Big)$.
 
\noindent The appearing fields $\psi_r(x)$, $\bar{\psi}_r (x)$, $\psi_r^c (x)= C \, \gamma_0 \,\psi_r^\ast (x)$, $\overline{\psi_r^c} (x)= \psi_r^T(x) \, C$ are (anti)causal Grassmann fields.\footnote{After {\em Canonical real-equal time quantization} the momentum-space annihilation and creation operators respect anticommutation relations (e.g.\ $\{b_r(\vec{p},s),d_{r^\prime}^+(\vec{p}^{\,\prime},s^{\,\prime})\}= (2\pi)^3 \,2\omega (\vec{p}\,)\,\delta^3(\vec{p}-\vec{p}^{\,\prime}) \, \delta_{s s^\prime}\delta_{r r^\prime}, \ldots$) containing again an {\em indefinite metric}.} The apparent complications with the construction of (anti)causal (non)Abelian vector fields have been discussed e.g.\ in \cite{Kleefeld:2003zj,Kleefeld:2002gw}.\footnote{The realization of such fields heavily relies on the existence of a renormalizable and unitary formalism for massive (non)Abelian vector fields without a Higgs mechanism provided by Jun-Chen Su \cite{Kleefeld:Su:1998wy}.} The construction of Dirac-spinors and polarization vectors for (anti)causal fields rely both on a formalism of Lorentz-boosts for complex mass fields \cite{Kleefeld:2003zj,Kleefeld:2002au,Kleefeld:2002gw,Kleefeld:2001xd} which for non-real masses show different group properties than the standard Lorentz boosts on the real mass shell.\footnote{As a consequence we had to note \cite{Kleefeld:2003zj} that (anti)causal fields are representations of the covering group of the complex Lorentz group $L_+(C)$ (or, more generally, the covering group of the respective Poincar\'{e} group) \cite{Kleefeld:Greenberg:2003nv}, while standard Hermitian (shadow) fields are representations of the covering group of the real Lorentz group $L^\uparrow_+$. The beauty of the group $L_+(C)$ is that it covers also non-Hermitian theories with a real spectrum and a probability interpretation like the PT-symmetric ones dicussed by C.M.\ Bender et al.\ \cite{Bender:2003ve}. 
G.P.\ Pron'ko \cite{Kleefeld:Pron'ko:1996be} could of course argue that such a Lorentz boost between $\vec{p}$ and $\vec{p}^{\;\prime}$ {\em ``$\ldots$ understood literary leads to nonsense because the transformed space components of the momentum become complex. $\ldots$''} Certainly this argument is {\em only} true for complex mass fields with finite $\Gamma$ being treated as {\em asymptotic states}. Yet --- as argued in the context of the Nakanishi-KG propagator ---  for complex mass fields in {\em intermediate states} Poincar\'{e} invariance is completely restored! Hence, we think the artificial construction of a Poincar\'{e} Semigroup by A.\ Bohm et al. \cite{Bohm:2002ja} is unnecessary and in strong conflict with the causal properties of asymptotic states.} In the footnote we conclude this section with a comment on the probability concept in (A)CQT.\footnote{As $|\psi(x)|^2$ is not a probability density in (anti)causal Schr\"odinger theory \cite{Kleefeld:2002au}, $|T_{fi}|^2$ is {\em not} to be interpreted as a {\em transition probability} in (anti)causal scattering theory! In (anti)causal scattering theory we have instead to consider a quantity $\overline{T}_{fi} \,T_{fi}$, where $\overline{T}_{fi}$ ($\not= T^+_{fi}$) is called the {\em conjugate $T$-matrix}. The construction of the explicit analytical expression for the conjugate causal T-matrix $\overline{T}_{fi}$ showed up to be a non-trivial task. As in Ref.\ \cite{Kleefeld:2003zj} we want to give here the final result without proof. We assume the causal $T$-matrix $T_{fi}$ to be determined by the standard expression in the interaction picture\\[1mm]
$(2\pi)^4 \, \delta^4 (P_f - P_i)\; i \; T_{fi} = $\\[1mm]
\makebox[5mm]{}$=\left<\!\left<0\right|\right. {\cal A}_{\,N^\prime_f} \ldots {\cal A}_{\,1^\prime} \; T [ \, \exp(i \, S_{int}) - 1 \, ]\; {\cal C}^+_1 \ldots {\cal C}^+_{N_i} \left|0\right>_c =: \left<\!\left<\psi_f\right|\right. T [ \, \exp(i \, S_{int}) - 1 \, ] \left|\psi_i\right>$\\[1mm]
with ${\cal A}_{j^\prime} \in \{ a (\vec{p}^{\,\,\prime}_{j}), b (\vec{p}^{\,\,\prime}_{j})\}$, ${\cal C}_j\in \{ c (\vec{p}_{j}), d (\vec{p}_{j})\}$. Call $N_F$ the overall number of Fermionic operators in the intitial and final state. Then we obtain for $\overline{T}_{if}$:\\
$(2\pi)^4 \delta^4 (P_f - P_i)\; (-i) \;\, \overline{T}_{if} =\left<\!\left<\bar{0}\right|\right. \Big( {\cal A}_{\,N_i} \ldots {\cal A}_{\,1} \; T [ \, \exp(- i \, S_{int}) - 1 \, ]\; {\cal C}^+_{\,1^\prime} \ldots {\cal C}^+_{\,N^\prime_f} \Big)^T \left|\bar{0}\right>_c$\\
\makebox[38.5mm]{}$= \; \left|\bar{0}\right>^T {\cal A}_{\,N_i} \ldots {\cal A}_{\,1} \; T [ \, \exp(- i \, S_{int}) - 1 \, ]\; {\cal C}^+_{\,1^\prime} \ldots {\cal C}^+_{\,N^\prime_f} \; \Big(\left<\!\left<\bar{0}\right|\right.\Big)^T $\\
\makebox[38.5mm]{}$=: \left|\right.\!\overline{\psi}_i\!\left.\right>^T \, T [ \, \exp(- i \, S_{int}) - 1 \, ]  \left(\left<\!\left<\right.\right.\!\overline{\psi}_f\!\!\left.\left.\right|\right.\right)^T $\\[2mm]
with ${\cal A}_{j} \in \{ a (\vec{p}_{j}), b (\vec{p}_{j})\}$, ${\cal C}_{j^\prime}\in \{ c (\vec{p}^{\,\,\prime}_{j}), d (\vec{p}^{\,\,\prime}_{\,j})\}$ and
$\left|\bar{0}\right>$ (and $\left<\!\left<\bar{0}\right|\right.$) being the {\em dual vacuum} annihilating creation operators and creating annihilation operators.
The transition probability for a causal process (being not necessarily a real number!) is given by\\[1mm]
$\overline{T}_{fi} \,T_{fi}= \left|\right.\!\overline{\psi}_i\!\left.\right>^T\, T [ \, \exp(- i \, S_{int}) - 1 \, ]  \left(\left<\!\left<\right.\right.\!\overline{\psi}_f\!\!\left.\left.\right|\right.\right)^T\! \left<\!\left<\psi_f\right|\right. T [ \, \exp(i \, S_{int}) - 1 \, ] \left|\psi_i\right>$\\[1mm]
\makebox[11mm]{}$=(-1)^{N_F(N_F-1)/2}\! \left<\!\left<\right.\right.\!\overline{\psi}_f\!\!\left.\left.\right|\right. \!\Big( T [ \, \exp(- i \, S_{int}) - 1 \, ]\Big)^T \! \left|\right.\!\overline{\psi}_i\!\left.\right> \left<\!\left<\psi_f\right|\right. T [ \, \exp(i \, S_{int}) - 1 \, ] \left|\psi_i\right>$.\\[1mm]
Hence the (complex) ``probability'' of a state $\left|\psi\right>$ to be in a state $\left|Y\right>$ appears to be $\left|\right.\!\overline{\psi}\!\left.\right>^T \! \left<\!\left<\right.\right.\!\overline{Y}\!\!\left.\left.\right|\right.^T \left<\!\left<Y\right|\right.\!\!\left.\psi\right>=
\left|\right.\!\overline{Y}\!\left.\right>^T \! \left<\!\left<\right.\right.\!\overline{\psi}\!\!\left.\left.\right|\right.^T
\left<\!\left<\psi\right|\right.\!\!\left.Y\right>$. For a further short discussion see Ref.\ \cite{Kleefeld:2003zj}.}

\section{(Anti)causal Quantum Mechanics}
The representation independent, time-dependent Schr\"odinger equation is $ i \, \partial_t \, \left|\psi(t)\right> = H \, \left|\psi(t)\right>$. Its adjoint is given by $- i\, \partial_t \left<\!\left<\psi(t) \right|\right. = \left<\!\left<\psi(t) \right|\right. H$. 
In 1-dim.\ QM we consider the Hamilton operator of the (anti)causal Harmonic Oscillator $H = H_C + H_A = \frac{1}{2} \, \omega \, [c^+, a\, ]_\pm + \frac{1}{2} \, \omega^\ast \, [a^+, c\, ]_\pm =  \omega \, (c^+ a \pm \frac{1}{2}) + \omega^\ast \, (a^+ c \pm \frac{1}{2})$ ($\pm$ for Bosons/Fermions\footnote{The Fermionic case we tend to denote by $H = \frac{1}{2} \, \omega \, \{\,d^+, b\, \} + \frac{1}{2} \, \omega^\ast \, \{\,b^+, d\, \}$.}) \cite{Kleefeld:Nakanishi:1972pt,Kleefeld:2003zj,Kleefeld:2002gw,Kleefeld:2001xd,Kleefeld:1998yj,Kleefeld:1998dg,Kleefeld:thesis1999,Kleefeld:Kossakowski2002} with\footnote{All futher commutators/anticommutators (Bosons/Fermions) of $a$, $c$, $a^+$, $c^+$ vanish!}\\[1mm]
\makebox[5mm]{}$\left( \begin{array}{cc} {[c,c^+]_\mp} & {[c,a^+]_\mp} \\
{[a,c^+]_\mp} & {[a,a^+]_\mp} \end{array}\right) \quad = \quad \left( \begin{array}{cc} 0 & 1 \\
1 & 0 \end{array}\right) \quad = \quad \mbox{``indefinite metric''}$ ,\\[1mm] 
yielding $[H_C,H_A]=0$ \footnote{A ``Hermitian'' Hamilton operator quantized with a non-trivial (e.g.\ indefinite) metric is called {\em pseudo-Hermitian}! Pseudo-Hermiticity (\& pseudo-unitarity) is presently promoted \cite{Kleefeld:Znojil2003} by M.\ Znojil and A.~Mostafazadeh. Ideas go back to names like W.\ Heisenberg, W.~Pauli, P.A.M.\ Dirac, S.N.\ Gupta, K.\ Bleuler, E.C.G.\ Sudarshan, K.L.\ Nagy.}. 
The (stationary, normalized) right eigenstates $\left|n,m\right>$ and left eigenstates $\left<\!\left<\,n,m\right|\right.$ for the eigenvalues $E_{n,m} = \omega  \, (n \pm \frac{1}{2}) + \omega^\ast \, (m \pm \frac{1}{2})$ are given by $\left|n,m\right> \; := \; (c^+)^n (a^+)^m \left|0\right>(n!\,m!)^{-1/2}$ and $\left<\!\left<\,n,m\right|\right. \; := \; \left<\!\left<\,0\right|\right. c^m \, a^n (m!\,n!)^{-1/2}$ (Bosons: $n,m\in\{0,1,2,\ldots\}$; Fermions: $n,m\in\{0,1\}$).
The (bi)orthogonal eigenstates are complete. In holomorphic representation (e.g.\ \cite{Kleefeld:Chruscinski2002}) the time-dependent Schr\"odinger equation and its adjoint are:\\[2mm]
\makebox[5mm]{} $+ \, i \, \partial_t \, \left<\!\left<z,z^\ast\right| \psi(t)\right>  =  \int dz^\prime dz^{\prime\ast} \, \left<\!\left<\right.\right.\!\!z,z^\ast | H |z^\prime,z^{\prime\ast}\!\left.\right> \, \left<\!\left<\right.\right.\!\!z^\prime,z^{\prime\ast}\left| \psi(t)\right>$,\\[2mm]
\makebox[5mm]{} $-\,i\, \partial_t \, \left<\!\left<\psi(t)\right| z,z^\ast\right>  =  \int dz^\prime dz^{\prime\ast} \,\left<\!\left<\right.\right.\!\!\psi(t)|z^\prime,z^{\prime\ast}\!\left.\right>\, \left<\!\left<\right.\right.\!\!z^\prime,z^{\prime\ast} | H\!\left|z,z^\ast\right>$.\\[2mm]
\noindent The holomorphic representation of $H$ is ($H(z,z^\ast)=H_C(z)+H_A(z^\ast)$) \cite{Kleefeld:2003zj}:\footnote{Note {\em different} ways to define a time-reversal operation  $(i^2=-1)$: e.g.\ $T:$ $z\rightarrow z$, $p\rightarrow -p$, $z^\ast\rightarrow z^\ast$, $p^\ast\rightarrow - p^\ast$, $i\rightarrow i$
 or ${\cal T}:$ $z\rightarrow z^\ast$, $p\rightarrow p^\ast$, $z^\ast\rightarrow z$, $p^\ast\rightarrow p$, $i\rightarrow -i$.}\\[2mm]
$H(z^\prime,z^{\prime\ast};z,z^\ast) = \left<\!\left<\right.\right.\!\!z^\prime,z^{\prime\ast} | H\!\left|z,z^\ast\right> = H(z,z^\ast) \left<\!\left<\right.\right.\!\!z^\prime,z^{\prime\ast} \! \left|z,z^\ast\right> =$ \\[1mm] 
\makebox[1mm]{}${\displaystyle = \Big( \! - \frac{1}{2\, M} \frac{d^2}{dz^2} + \frac{1}{2} \; M \, \omega^2 \, z^2  \, - \frac{1}{2\, M^\ast} \frac{d^2}{dz^{\ast\,2}} + \frac{1}{2} \; M^\ast \, \omega^{\ast \,2} \, z^{\ast\,2} \Big) \left<\!\left<\right.\right.\!\!z^\prime,z^{\prime\ast} \!\left|z,z^\ast\right>}$.

\acknowledgements

This work has been supported by the
{\em Funda\c{c}\~{a}o para a Ci\^{e}ncia e a Tecnologia} \/(FCT) of the {\em Minist\'{e}rio da Ci\^{e}ncia e da Tecnologia (e do Ensinio Superior)} \/of Portugal, under Grants no.\ PRAXIS
XXI/BPD/20186/99, SFRH/BDP/9480/2002, and POCTI/\-FNU/\-49555/\-2002.

\end{document}